\documentclass[
 preprint,footinbib,
 amsmath,amssymb,
 aps,prb,showpacs, showkeys
]{revtex4-1}
\usepackage[utf8]{inputenc}
\usepackage{graphicx}
\DeclareGraphicsExtensions{.png}
\usepackage{dcolumn}
\usepackage{bm}
\usepackage{url}
\usepackage{amsmath}
\let\oldAA\AA
\renewcommand{\AA}{\text{\normalfont\oldAA}}

\begin{document}
\title{Effects of Ba doping on the structural and electronic properties of La$_{2-x}$Ba$_x$CuO$_4$}
\author{E. Y. Soto-Gómez}
\affiliation{Grupo de Investigación en Ciencias Básicas, Aplicación e Innovación - CIBAIN, UNITR\'OPICO, Yopal, Casanare, Colombia}

\author{G. I. Gonz\'alez-Pedreros}
\affiliation{Departamento de F\'isica, Facultad de Ciencias, Universidad Militar Nueva Granada, Bogotá, Colombia}

\author{J. A. Camargo-Mart\'inez}
\email{jcamargo@unitropico.edu.co}
\affiliation{Grupo de Investigación en Ciencias Básicas, Aplicación e Innovación - CIBAIN, UNITR\'OPICO, Yopal, Casanare, Colombia}

\date{\today}

\begin{abstract}

The structural and electronic properties of La$_{2-x}$Ba$_x$CuO$_4$ were investigated as a function of Ba-concentration $0 \leq x \leq 1.2$, within the virtual crystal approximation (VCA) by means of first-principles total-energy calculations. We found Ba doping induces a quasi-rigid displacement of Cu d$_{x^{2}-y^{2}}$ and d$_{z^{2}}$ bands toward higher energies. This effect generates important changes in the Fermi surface topology, which manage to imitate the ARPES momentum distribution map (MDM) measurements and tight-binding calculations recently reported for doped La-based cuprates. The calculations exhibit the significant increase in the contribution of Cu d$_{z^{2}}$ states at the Fermi level for greater Ba doping content to 0.3, which has effects on orbital hybridisation with Cu d$_{x^{2}-y^{2}}$ states and therefore in the suppression of the superconducting state. For this Ba doping range, no magnetism was found.

\end{abstract}

\maketitle

\section{Introduction}

Recently, important experimental results of electronic properties of La-based cuprates have been reported~\cite{A,Ax1,Ax2,Ax3}. One of them provided direct ultraviolet and soft-X-ray angle-resolved photoelectron spectroscopy (ARPES) measurements of the orbital hybridisation between d$_{x^{2}-y^{2}}$ and d$_{z^{2}}$ states~\cite{Ax1}. This interaction plays a sabotaging role for superconductivity in this kind of systems~\cite{Ax1}. It has been shown that the hybridization of d$_{z^{2}}$ states on the Fermi surface substantially affects T$_c$ in the cuprates~\cite{AA2}. These results also provide an explanation for the Fermi surface (FS) topology and the proximity of the van-Hove singularity to the Fermi level. Both experimental and theoretical studies about the FS topology and its relationship with superconductivity have become an interesting research topic. 
In order to understand the mechanisms that control of physical properties of high-Tc cuprate systems, it is necessary to determine and study their structural and electronic properties both experimentally and theoretically. La$_{2-x}$M$_x$CuO$_4$ (M = Ba, Sr) layered perovskites have a strongly dependent on the hole concentration~\cite{A}, therefore is necessary to investigate in order to extract key features relevant of their normal and superconducting state properties. Many investigations have focused on underdoped regime~\cite{5xx,6xx} but little attention has been devoted to the overdoped regime, where the abrupt suppression of superconductivity for critical doping cannot be easily understood in terms of the standard BCS theory~\cite{7xx}.
One role of Ba in La$_{2-x}$Ba$_x$CuO$_4$ is to stabilize the higher-symmetry tetragonal {\it I4/mmm} phase, which is metallic and superconducting in the Ba doping range of 0.05 to 0.3~\cite{1,2,4,5}, with a maximum superconducting transition temperature (T$_c^{max}$) of 30 K for $x\sim 0.15$~\cite{5,3,13}. The primary interest in this material is in understanding what interaction between electrons is responsible for producing the superconducting state~\cite{pic}.
In this work, we report a detailed study of the effects the Ba doping on the structural and electronic properties of La$_{2-x}$Ba$_x$CuO$_4$ for underdoped and overdoped regimes ($0 \leq x \leq 1.2$), using the ab initio virtual crystal approximation (VCA). The evolution of the electronic band structure, density of states (DOS) and 2D-3D Fermi surface (FS) as a function of Ba doping is analyzed and compared with recent experimental and theoretical results.

\section{Method of Calculation}

To calculate the structural and electronic properties of La$_{2-x}$Ba$_x$CuO$_4$ we used the Full-Potential Linearized Augmented Plane Wave Method (LAWP)~\cite{9} within the local density approximation (LDA), as implemented in the WIEN2K code~\cite{9a}. As a cut-off for base functions we used $R_{MT}K_{MAX}=$ 8.0 and for the expansion of the functions of the base within the atomic spheres a maximum angular momentum value $l_{MAX}=12$ with $G_{MAX}=25$ and a $14\times14\times14$ grid on the k-space which contains 240 points in the irreducible of the Brillouin zone (IBZ). The muffin-tin sphere radii, Rmt (in atomic units), were 2.15 for La, 1.72 for Cu and 1.48 for O. For the substitution of Ba for La, we used the Virtual Crystal Approximation (VCA)~\cite{B,C}. The La ($Z=57$) sites are substituted for pseudo-atoms which have a fractional electronic charge ($Z=57-x$), depending on the Ba concentration $x$. This approximation is justified mainly by the fact that La only has one electron more than Ba.
Our calculations were done for Ba concentrations $x = $ 0, 0.03, 0.05, 0.075, 0.1, 0.125, 0.15, 0.20, 0.3, 0.40, 0.5, 0.60, 0.80, 1.0 and 1.2.

\section{Results and discussion} 

\subsection{Structural properties}

First, we studied the Ba doping effects on the structural properties of La$_{2-x}$Ba$_x$CuO$_4$ with a body-centered-tetragonal structure (bct) and space group $I4/mmm$~\cite{1,2}. The crystal structure consists of weakly coupled twodimensional CuO planes, separated by the LaO buffers~\cite{D}. In this work, we define the oxygen of the Cu planes as O1 and the apical oxygen as O2.  

\begin{figure}[!ht]
\centering
\includegraphics[width=0.65\textwidth]{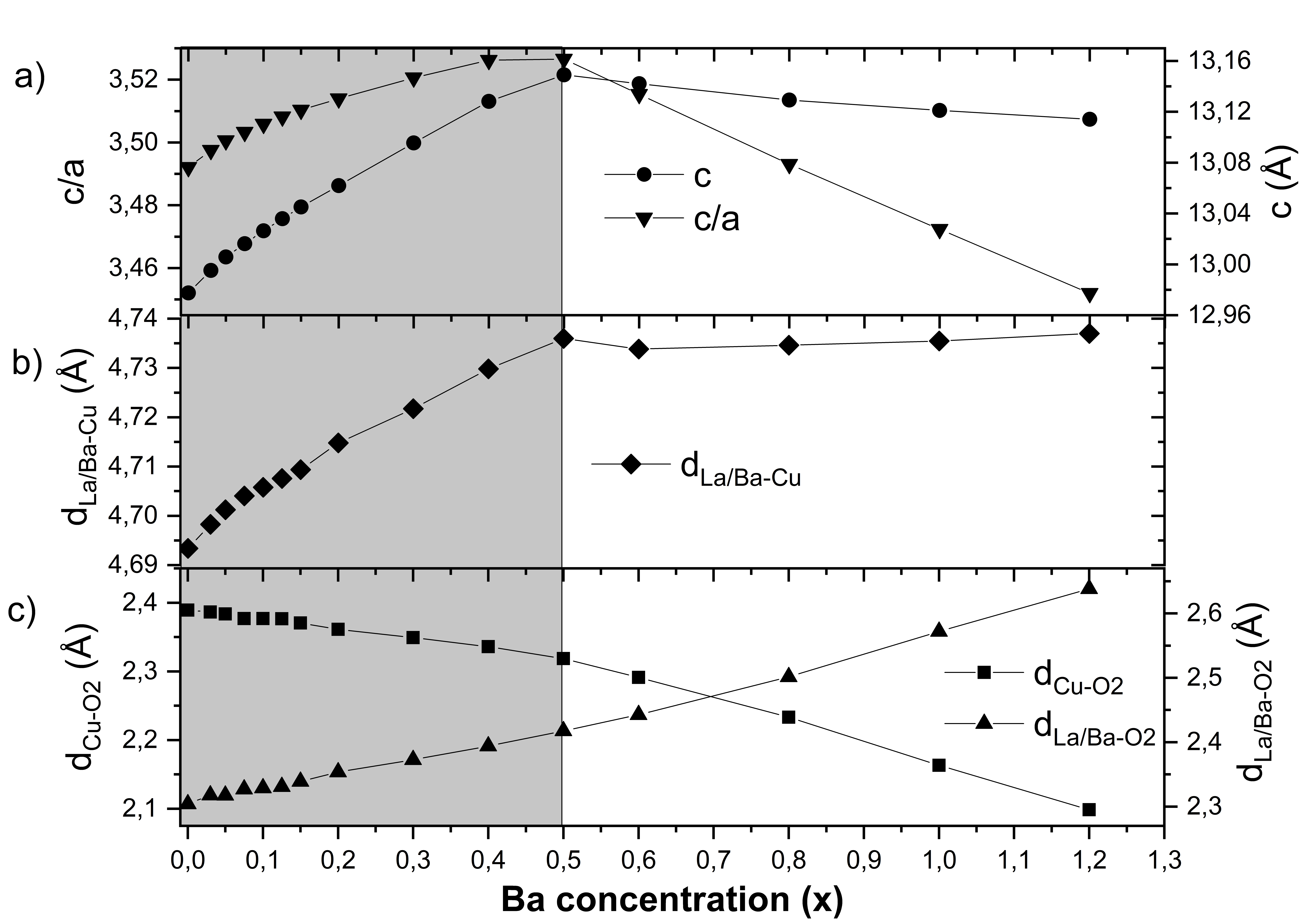}
\caption{a) Lattice parameter {\it c} and {\it c/}a ratio. b) La/Ba-Cu (d$_{La/Ba-Cu}$), c) Cu-O2 (d$_{Cu-O2}$) and La/Ba-O2 (d$_{La/Ba-O2}$) relative distances as a function of Ba doping in the La$_{2-x}$Ba$_x$CuO$_4$ compound.}
\label{F1} 
\end{figure}

Starting from the experimental parameters of La$_{1.85}$Ba$_{0.15}$CuO$_4$ measured at 10 K~\cite{2}, the {\it c/a} ratio was optimized by minimizing the total energy at a constant volume and the atomic coordinates were relaxed by minimizing the total force for each Ba concentration. We show in Fig.~\ref{F1} the effects of Ba doping on the lattice parameter {\it c}, the {\it c/a} ratio, La/Ba-Cu (d$_{La/Ba-Cu}$), Cu-O2 (d$_{Cu-O2}$) and La/Ba-O2 (d$_{La/Ba-O2}$) relative distances. 

It is observed that Ba doping apparently induces in the structural properties two different behaviors, identified as gray and white regions in Fig.~\ref{F1}. Below a Ba doping of 0.5 (gray region) the structure shows a significant increase in the z-axis (see Fig.~\ref{F1}a), while the lattice parameter {\bf a} is almost constant ($\approx3.715$ \AA). This behavior is in good agreement with the experimental one~\cite{nini}, with errors below 2\% ($\sim 2 \AA $). For greater doping content to 0.5 (white region), the parameter {\it c} remains almost constant. The structure presents a certain degree of stabilization in the z-direction while the {\it c/}a ratio decreases, e.g., an increase in the area of {\it xy} plane. The Fig.~\ref{F1}.b show that the doping below 0.5 induces a progressive increase in the d$_{La/Ba-Cu}$, while remains constant for great doping content to 0.5 suggesting structural stability in the system, with a behavior similar to the
one observed in the parameter {\it c}.

The behavior of the d$_{Cu-O2}$ y d$_{La/Ba-O2}$ observed in the Fig.~\ref{F1}c differs of d$_{La/Ba-Cu}$, which is associated to the displacement of the O2 atom between the Cu and La/Ba planes by the increase of the Ba concentrations. This behavior can be explained by Coulomb interaction between the Cu and La/Ba planes. The ionic character of the La/Ba planes in the crystal structure tends to attract electrons towards La/Ba plane competing with the electronic affinity of Cu-O1 plane. The increase of the Ba concentrations changes the ionic character of the La/Ba planes and as a consequence, the O2 atom is attracted easily to the Cu plane. So, less energy is needed to promote the Cu 3d hole into the $z^2$ orbital pointing toward apical O~\cite{Ho}. The O2 atom acts as a bridge for the charge transfer to La/BaO buffers, a similar behavior was observed in Bi2223~\cite{10,10a}.

\subsection{Electronic properties}

\begin{figure}[!ht]
\centering
\includegraphics[width=1\textwidth]{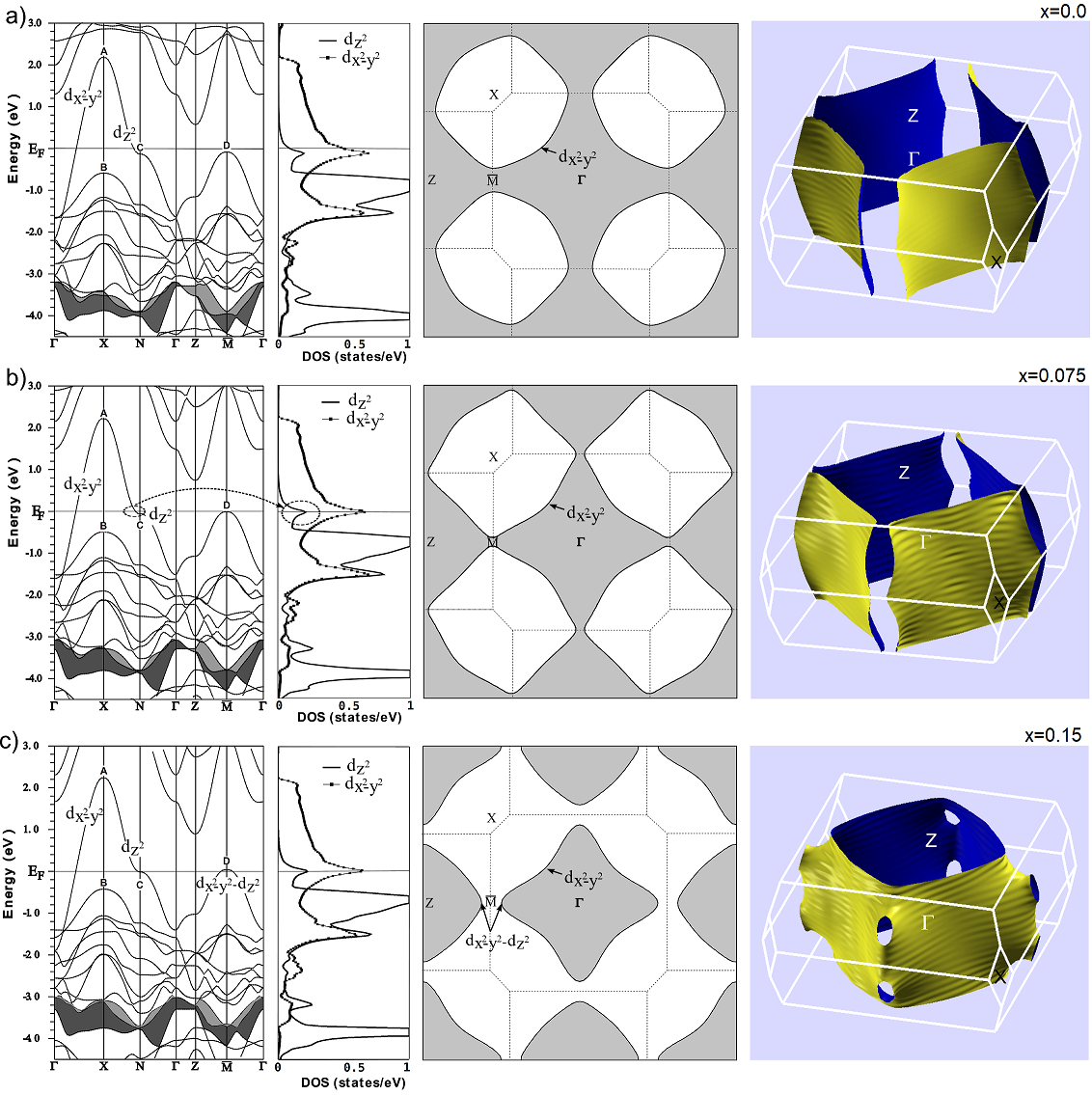} 
\caption{ (color online) Electronic band structure, projected density of states (PDOS) and Fermi surface (FS) ($k_z=0$ plane and 3D) of La$_{2-x}$Ba$_x$CuO$_4$ for Ba concentrations of a) $x=0$ (undoped), b) 0.075 and c) 0.15.}
\label{F2} 
\end{figure}

\begin{figure}[!ht]
\centering
\includegraphics[width=1\textwidth]{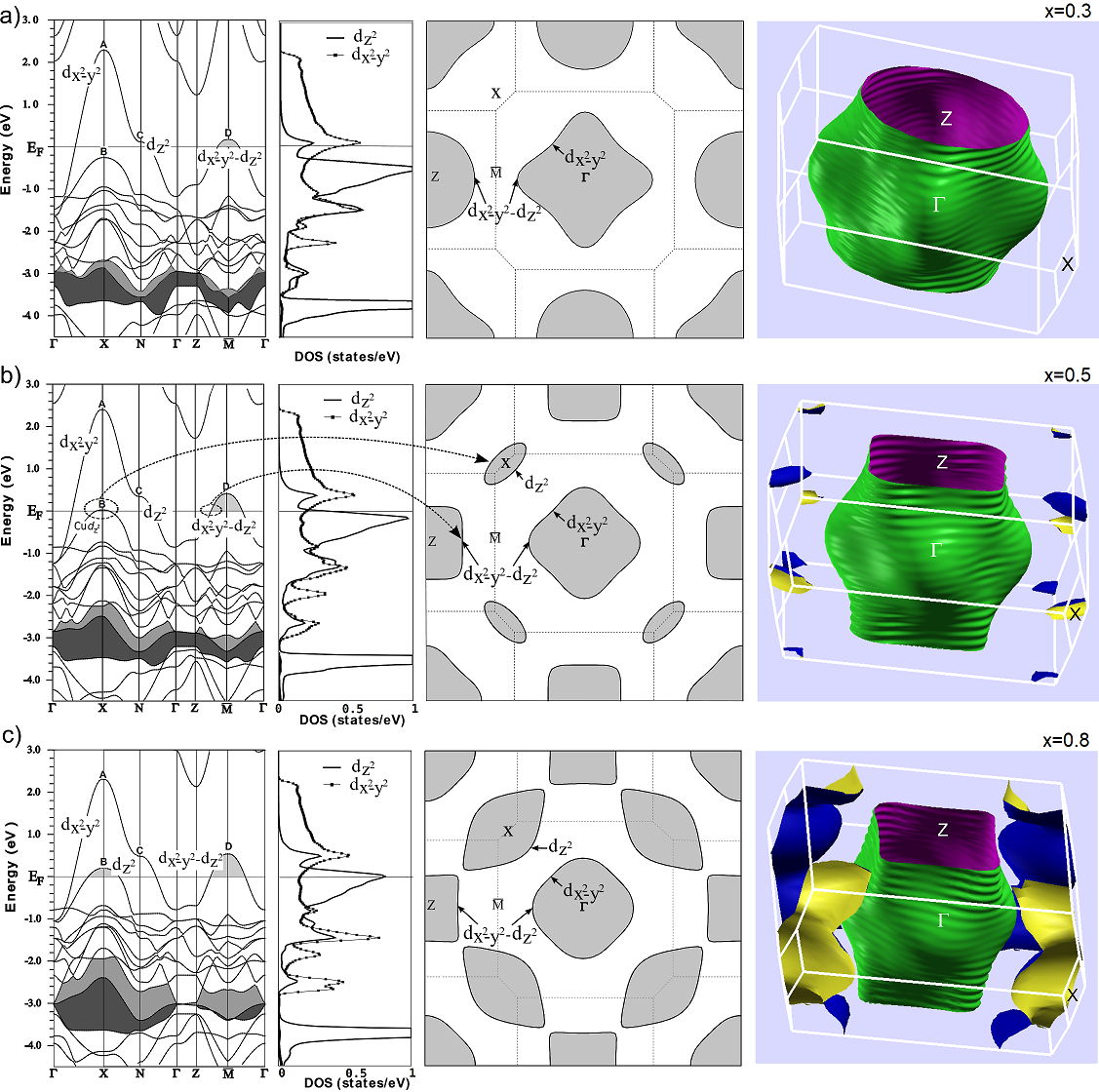} 
\caption{(color online) Electronic band structure, projected density of states (PDOS) and Fermi surface (FS) ($k_z=0$ plane and 3D) of La$_{2-x}$Ba$_x$CuO$_4$ for Ba concentrations of a) 0.3, b) 0.5 and c) 0.8.}
\label{F3} 
\end{figure}

The effects of Ba doping on the electronic band structures, projected density of states (PDOS) and Fermi surfaces (FS) of La$_{2-x}$Ba$_x$CuO$_4$ for concentrations of $x=0$ (undoped), 0.075 and 0.15 are shown in Fig.~\ref{F2}, and for concentrations of 0.3, 0.5 and 0.8 in Fig.~\ref{F3}, respectively.

The electronic properties calculated for undoped La$_{2}$CuO$_4$ reveal the typical contributions at Fermi level (E$_F$) from Cu 3d$_{x^{2}-y^{2}}$ - O1 2p$_{xy}$ states with a small contribution of Cu 3d$_{z^{2}}$ - O2 2p$_{z}$ states ($pd\sigma$) generating a closed Fermi surface around X point in IBZ (see Fig.~\ref{F2}a), in a well agreement with calculations reported in the literature~\cite{3,11}. Recently Matt et al.~\cite{Ax1} reported for first time angle-resolved photoelectron spectroscopy (ARPES) measurements on La-based cuprates that provided direct observation of this states, which confirm theoretical results.

In general, when Ba is introduced in La$_{2}$CuO$_4$, it induces a quasi-rigid displacement of the Cu d$_{x^{2}-y^{2}}$ and Cu d$_{z^{2}}$ states toward higher energies (see band structures and PDOS in Fig.~\ref{F2} and Fig.~\ref{F3}), generating important changes in the contributions at E$_F$, which have been observed experimentally by ARPES for La$_{2-x}$Sr$_x$CuO$_4$ (LSCO)~\cite{A,Ax1,Ax2}.

In Fig.~\ref{F2}b it is observed in the PDOS for 0.075 Ba doping the contribution at E$_F$ of two peaks, which correspond to Cu d$_{x^{2}-y^{2}}$ and Cu d$_{z^{2}}$ states (near to N point), these contributions are associated to energy bands in $\Gamma$-X-N path. These peaks are related with the extended Van Hove Singularities (VHS), which play an important role in the physics of high-temperature superconductors~\cite{14,15}. The superconducting state in La$_{2-x}$Ba$_x$CuO$_4$ for Ba doping between 0.05 and 0.3 has been experimentally observed, with a T$_c^{max}$ of 30 K at $x\sim 0.15$~\cite{3,5,13}.  

An important change occurs for greater doping to 0.1 where contributions of Cu d$_{x^{2}-y^{2}}$-d$_{z^{2}}$ states at E$_F$ around $\bar{M}$ point are observed. The FS calculated to $x=0,075$ concentration (see Fig.~\ref{F2}b) reveals concentric closed surfaces of Cu d$_{x^{2}-y^{2}}$ states around X point of the IBZ which are noticeably flattened in the X–Z direction. These contributions generate a seemingly well nested FS (square shape in X). The FS near $\bar{M}$ softly moves through $\bar{M}$ so that the topological center of the FS is turned over from X to $\Gamma$ since $x\sim 0.1$ and additionally a closed surface is generated around Z (see FS in Fig.~\ref{F2}c). The Fig.~\ref{F2} also show the significant variations of 3D FS for Ba concentrations of 0, 0.075 and 0.15.

The dispersion bands between 3 and 4 eV below E$_F$ are mainly due to Cu d$_{z^{2}}$ states which hybridize with Cu d$_{xy}$ and Cu d$_{xz}$ states (Fig.~\ref{F2}). Ba doping of $x=0, 0.075$ and 0.15 no have important effects in this energy range.

Now, for 0.3 Ba doping the area of the closed-hole surface around $\Gamma$ and Z points decreased and important changes in the 3D FS are observed (see Fig.~\ref{F3}a). For Ba concentrations between 0.5 and 1.2, it is observed a new concentric closed surface around X point associated to Cu d$_{z^{2}}$ states, whose area increases with the increase of doping (Fig.~\ref{F3}b and c). For 0.8 Ba doping a seemingly well nested FS round Z and $\Gamma$ points are observed (Fig.~\ref{F3}c). Regarding the band structure for Ba doping of 0.3, 0.5 and 0.8, we observe a significant increase in the shaded area (at 3-4 eV below E$_F$), which is associated with changes in the hybridisation of Cu d$_{z^{2}}$ with Cu d$_{xy}$ and Cu d$_{xz}$ states.

\begin{figure}[!ht]
\centering
\includegraphics[width=0.47\textwidth]{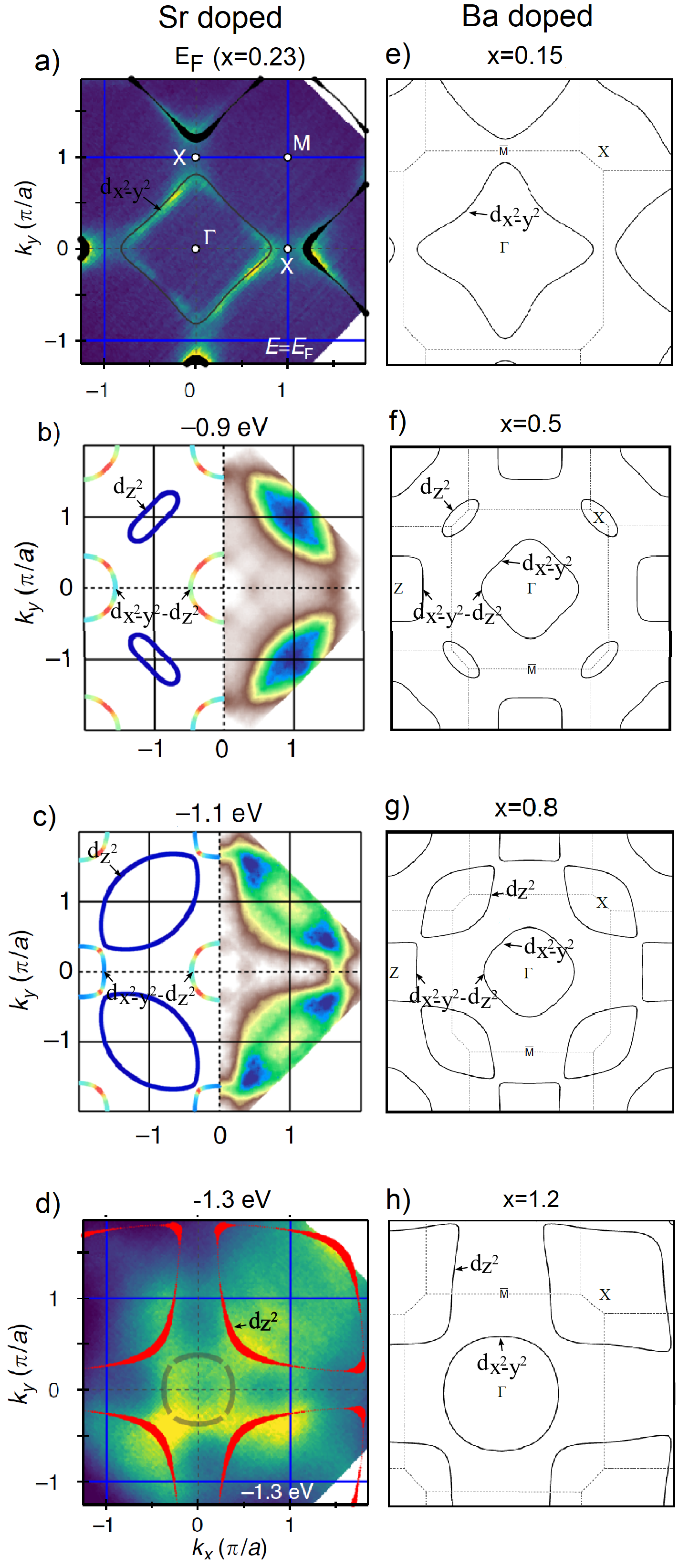} 
\caption{Analogy in the Fermi Surface behavior of raw ARPES momentum distribution maps (MDM) of La$_{2-x}$Sr$_x$CuO$_4$ for $x = 0.23$ at different binding energies and the FS's calculated in this work. 
In a) and d) MDM reported by Matt et al.~\cite{Ax1}, b) and c) MDM reported by Horio et al.~\cite{Ax2}. b) and c) also show tight-binding surfaces calculated by Horio et al.~\cite{Ax2}. In e) - h) our FS calculated ($x=0.15,0.5,0.8,1.2$). Position of Cu d$_{x^{2}-y^{2}}$ and Cu d$_{z^{2}}$ bands are shown in each case. Note the different notation for IBZ points in a).}
\label{F4} 
\end{figure}

In general, it is observed in FS (Fig.~\ref{F2} and ~\ref{F3}) that the hole concentration (area) around $\Gamma$ decreases with the increase of doping, which is consistent with the Luttinger sum rule for the electron density 1-x~\cite{A}, as well as the increase of the area of the X point. It is important to note that d-state contributions in half $\Gamma$-X direction are a little sensitive to doping with Ba. The hybridisation between Cu d$_{x^{2}-y^{2}}$ and d$_{z^{2}}$ orbitals it is observed in the $\Gamma$-$\bar{M}$-Z direction.

In Fig.~\ref{F4} it is presented the analogy in the behavior in the Fermi Surface behavior of ARPES momentum distribution maps (MDM) of La$_{2-x}$Sr$_x$CuO$_4$ for $x = 0.23$ at different binding energies~\cite{Ax1,Ax2} and FS's calculated (Ba-doped) in this work, including the tight-binding SF calculated (Sr-doped) by Horio et al.~\cite{Ax2}. We observed that Ba doped procedure with VCA ($x \geq 0.15$) manage to mimic the behavior (both in shape and orbital character) both experimental and theoretical results (Sr doped) reported in the literature~\cite{Ax1,Ax2}. We observed that Ba-doped concentration of 0.15 in La$_{2-x}$Ba$_x$CuO$_4$ it is equivalent to Sr-doped concentration of 0.23 in La$_{2-x}$Sr$_x$CuO$_4$. Taking into account the total number of electrons of each atom (Z$_{Ba}=56$ and Z$_{Sr}={56}$), these concentrations (0.15$_{Ba}$ and 0.23$_{Sr}$) represent a similar hole-doped concentration (8.4 and 8.7, respectively), which justifies our analogy. Ba doping concentrations of $x=0.5,0.8$ and 1.2 induce a displacement of Cu d$_{z^{2}}$ band toward higher energies around X point (as can be seen in Figs.~\ref{F2} and ~\ref{F3}), which leads to the imitation of the ARPES momentum distribution map measurements in Sr-doped La$_2$CuO$_4$ at 0.9, 1.1 and 1.3 eV below $E_F$, respectively (see Fig.~\ref{F4}). This allows us to infer that our Ba-doped calculations for $x \geq 0.15$ of the electronic properties using VCA have an acceptable coherence with the electronic behavior of the system under study. Furthermore, our FS calculations around $\Gamma$ point also show similarity in orbital hybridisation (d$_{x^{2}-y^{2}}$ and d$_{z^{2}}$ states) with tight-binding surfaces calculated by Horio et al.~\cite{Ax2}.

\begin{figure}[!h]
\centering
\includegraphics[width=0.8\textwidth]{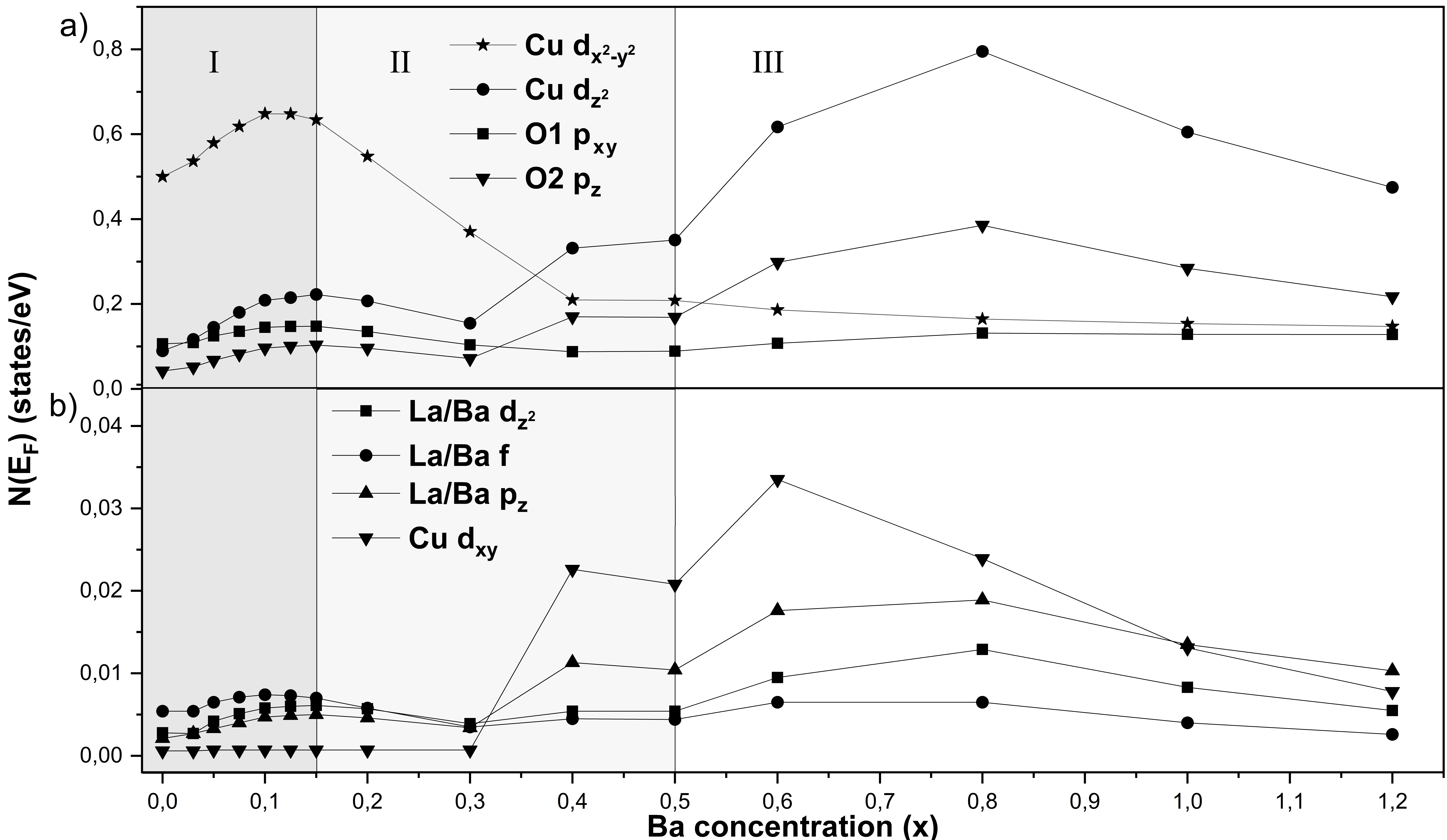} 
\caption{DOS at E$_F$, N(E$_F$), as a function of Ba doping in La$_{2-x}$Ba$_x$CuO$_4$ (a) DOS with higher contribution to N (E$_F$) y (b) DOS with less contribution to N (E$_F$). Note the differences in the scale of the y-axes.}
\label{F5} 
\end{figure}

Fig.~\ref{F5} shows Ba doping effects on the DOS at E$_F$, N(E$_F$), for La$_{2-x}$Ba$_x$CuO$_4$. The Cu d$_{x^{2}-y^{2}}$, Cu d$_{z^{2}}$, O1 p$_{xy}$ and O2 p$_{z}$ states are the dominant contributions at the N(E$_F$) (see Fig.~\ref{F5}a), while small contributions are due to Cu d$_{xy}$, La/Ba p$_{z}$, La/Ba f and La/Ba d$_{z^{2}}$ (Fig.~\ref{F5}b). The behavior of N(E$_F$) as an effect of Ba doping reveals three regions, labeled as I, II  and III in Fig.~\ref{F5}. 

Below Ba doping of 0.15 (region I) the main contribution is from Cu d$_{x^{2}-y^{2}}$ states (see Fig. 4.a), reaching a maximum N(E$_F$) at $x=0.10$ (6.5 states/cell). This value is related with T$_c$ at the optimal doping (T$_c^{max}$) measured in La$_{1.85}$Ba$_{0.15}$CuO$_4$~\cite{3,5}, confirming that a high N(E$_F$) is good for superconductivity~\cite{AA1}. However, at $x=0.12$ a slight decrease in the contribution of Cu d$_{x^{2}-y^{2}}$ states at E$_F$ begins. This Ba-doped concentration ($x\sim 1/8$) is known experimentally for an anomalous suppression of superconductivity~\cite{Mood,Kuma}, which seems to be related to spatial modulations of spin and charge density in CuO$_2$ planes with antiferromagnetic stacking~\cite{Tran}. To overcome this issue, an antiferromagnetic scenario is needed, which was not included in our calculations. In region II (0.15 $\leq x \leq 0.5$) it is observed the decrease in the contribution of Cu d$_{x^{2}-y^{2}}$ states, and the increase of the contribution of Cu d$_{z^{2}}$ and O p$_{z}$ states at E$_F$. Experimentally for greater doping content to 0.3 it is observed the suppression of superconductivity in La$_{2-x}$Ba$_x$CuO$_4$~\cite{3,5,13}, which occurs by the action of the hybridisation between d$_{x^{2}-y^{2}}$ and d$_{z^{2}}$ orbitals~\cite{Ax1,AA2}. Our results show that for this Ba doping (0.3) start the increase of the significant contribution of Cu d$_{z^{2}}$ and O p$_{z}$ states at E$_F$ and the hybridisation between d$_{x^{2}-y^{2}}$ and d$_{z^{2}}$ which, in a certain way, confirms the experimental and theoretical behavior reported. Finally, in region III it is observed the higher contribution of Cu d$_{z^{2}}$ and p$_{z}$ states at $x=0.8$ (0.8 states/eV) while Cu d$_{x^{2}-y^{2}}$ contribution remains almost constant with the increase in doping concentration. 

On the other hand, with respect to the lower contributions at E$_F$ (Fig.~\ref{F5}b), it is observed an appreciable effect of Ba doping on Cu d$_{xy}$ states for greater doping content to 0.3, with a maximum value (0.033 states/eV) at 0.6. These small contributions are due to a slight interaction between d$_{xy}$ and states d$_{z^{2}}$ around X point in IBZ.

For this Ba doping range, no magnetism was found. The system is below the Stoner limit for ferromagnetism, when the Ba concentration becomes larger ($x\geq 1.5$), ferromagnetism appears in VCA calculations~\cite{7xx}.

\section{Conclusions}

We have performed a first-principles study of the effects of Ba doping on the structural parameters, band structures, density of states (DOS) and 2D-3D Fermi surfaces (FS) of La$_{2-x}$Ba$_x$CuO$_4$ ($0 \leq x \leq 1.2$), using the virtual crystal approximation (VCA). In general, our results are correlated closely with both experimental and theoretical superconducting behavior reported for Ba-doped La${2}$CuO$_4$.

The band structures and density of states show that Ba doping induces a quasi-rigid displacement of Cu d$_{x^{2}-y^{2}}$ and d$_{z^{2}}$ bands toward higher energies, generating important changes in the Fermi surface (FS) topology. The FS calculations manage to imitate both shape and orbital character the ARPES momentum distribution map measurements and tight-binding calculations recently reported for doped La-based cuprates.

The hybridisation between Cu d$_{x^{2}-y^{2}}$ and d$_{z^{2}}$ orbitals was observed in the $\Gamma$-$\bar{M}$-Z direction. For greater Ba doping content to 0.3, the contribution of Cu d$_{z^{2}}$ states at the Fermi level ($E_F$) increases significantly, while the contribution of Cu d$_{x^{2}-y^{2}}$ states decreases. We found that both behaviors are consistent with the suppression of the superconducting state experimentally observed in La-based cuprates. The higher contribution of Cu d$_{x^{2}-y^{2}}$ states at $E_F$ for Ba doping concentrations between 0.125 and 0.15 correlates closely with the maximum superconducting critical temperature (30 K), measured in La$_{1.85}$Ba$_{0.15}$CuO$_4$, confirming that a high N(E$_F$) is good for superconductivity.

These results show that the VCA achieves a good reproduction of the electronic properties of Ba-doped La${2}$CuO$_4$.

\end{document}